\documentclass[english, 11pt]{article}
\usepackage{times}
\usepackage{cite}
\usepackage[latin9]{inputenc}
\usepackage{geometry}
\usepackage{textcomp}
\usepackage{babel}
\usepackage{array}
\usepackage{verbatim}
\usepackage{amsmath}
\usepackage{amssymb}
 \usepackage{algorithm}
 \usepackage{algorithmic}
\usepackage{graphicx}
\usepackage[unicode=true,pdfusetitle,
 bookmarks=true,bookmarksnumbered=false,bookmarksopen=false,
 breaklinks=false,pdfborder={0 0 1},backref=section,colorlinks=false]
 {hyperref}
\usepackage{breakurl}
\makeatletter

\usepackage{amsthm}\usepackage{dsfont}\usepackage{array}\usepackage{mathrsfs}\usepackage{comment}\onecolumn

\usepackage{color}
\usepackage{babel}

\let\bs=\boldsymbol

\def \saliency {\textup{\saliency}}

\def \path {\mathit{path}}

\def \label {\mathit{label}}

\usepackage{babel}
\usepackage{arydshln}
\date{}

\makeatother

\theoremstyle{plain}

\newtheorem{prop}{\textbf{Proposition}}

 \theoremstyle{definition}
\numberwithin{theorem}{section}
\numberwithin{lem}{section}

\theoremstyle{remark}

  \let\set=\mathcal \global\long\def\para#1{\noindent{\bf #1}}

\author{
Hashemifar, Somaye\\
  \texttt{hashemifar@ttic.edu}
  \and
  Huang, Qixing\\
  \texttt{huangqx@ttic.edu}
  \and
  Xu, Jinbo\\
  \texttt{j3xu@ttic.edu}\\ 
}

\title{Joint alignment of multiple protein-protein interaction networks via convex optimization}

\begin{document}

\maketitle
\noindent
\textbf{Abstract.}

\noindent
\textbf{Motivation:} High-throughput experimental techniques have been producing more and more protein-protein interaction (PPI) data. PPI network alignment greatly benefits the understanding of evolutionary relationship among species, helps identify conserved sub-networks and provides extra information for functional annotations. 
%With increasing availability of PPI networks of different species, it is gets more complicated to assess a biologically meaningful alignment in efficient time. 
Although a few methods have been developed for multiple PPI network alignment, the alignment quality is still far away from perfect and thus, new network alignment methods are needed.
\noindent \\
\textbf{Result:}  In this paper, we present a novel method, denoted as ConvexAlign, for joint alignment of multiple PPI networks by convex optimization of a scoring function composed of sequence similarity, topological score and interaction conservation score. In contrast to existing methods that generate multiple alignments in a greedy or progressive  manner, our convex method optimizes alignments globally and enforces consistency among all pairwise alignments, resulting in much better alignment quality. 
Tested on both synthetic and real data, our experimental results show that ConvexAlign outperforms several popular methods in producing functionally coherent alignments. ConvexAlign even has a larger advantage over the others in aligning real PPI networks.
ConvexAlign also finds a few conserved complexes among 5 species which cannot be detected by the other methods. 
%\input{main_text.tex}

%\liographystyle{IEEEtran}
%\bibliography{trans_sync_bib}

\section{Introduction}
Protein-protein interaction (PPI) networks provide valuable information for understanding of protein functions and system-level cellular processes. The alignment of PPI networks is a useful means for comparing the networks of different species. This comparison helps identify evolutionarily conserved pathways/complexes that may be functionally significant. Studying the conserved modules may provide useful information about the molecular mechanism contributing to their functions.

PPI networks can be aligned either locally or globally. Local network alignment methods such as  Mawish~\cite{koyuturk2006pairwise} and AlignNemo~\cite{ciriello2012alignnemo} aim to find small isomorphic subnetworks. 
%These methods search for conserved subnetworks using an alignment graph, in which nodes correspond to groups of orthologs and edges correspond to conserved interactins.  
Global network alignment (GNA) methods maximize the overall match between input networks. Some GNA methods such as IsoRank~\cite{singh2007pairwise, singh2008global}, MI-GRAAL\cite{kuchaiev2011integrative}, GHOST~\cite{patro2012global}, MAGNA~\cite{saraph2014magna, vijayan2015magna++}, Prob\cite{todor2013probabilistic}, NETAL~\cite{neyshabur2013netal} and HubAlign~\cite{hashemifar2014hubalign} are designed for pairwise alignment,  while others such as IsoRankN\cite{liao2009isorankn} and NetCoffee~\cite{hu2013netcoffee} for multiple alignment. GNA can be one-to-one or many-to-many mapping. The latter allows one protein to be aligned to multiple proteins of a single network while the former does not. 

More attention has been paid to pairwise network alignment. With the availability of more PPI networks, it becomes inevitable to align multiple networks. Existing GNA methods such as NetworkBlast-M~\cite{kalaev2008networkblast, sharan2005conserved} and GraemLin 2.0~\cite{flannick2008automatic} are designed for local alignment of multiple networks, whereas others such as IsoRankN~\cite{liao2009isorankn}, SMETANA~\cite{sahraeian2013smetana}, NetCoffee~\cite{hu2013netcoffee}, BEAMS~\cite{alkan2014beams} and FUSE~\cite{fuse2014Natasha} for global alignment of multiple networks. In addition to sequence similarity, all these methods excluding NetworkBlast-M and NetCoffee also employs topological information. Moreover, all the methods except NetCoffee are designed for many-to-many alignments. NetworkBlast-M starts with a set of highly conserved regions and then extends them greedily. GraemLin2.0 integrates phylogenetic information and network topology and then employs a hill-climbing algorithm to generate the alignment. IsoRankN applies IsoRank to compute the alignment scores between each pair of networks and then uses a PageRank-Nibble algorithm to cluster the proteins. SMETANA employs a semi-Markov random walk model to measure similarity between proteins. BEAMS constructs a weighted k-partite graph in which edges are assigned weights derived from protein sequence similarity. NetCoffee applies a triplet approach similar to T-Coffee to compute the edge weights of the k-partite graph. Both BEAMS and NetCoffee apply a heuristic on the k-partite graph to build an alignment. BEAMS fulfills this by greedily merging a set of disjoint cliques while NetCoffee by applying a simulated annealing method on a set of candidates. FUSE applies a non-negative matrix tri-factorization method to compute edge weights of the k-partite graph. 

Most of existing GNA methods do not optimize alignment of all proteins simultaneously. Instead, they start from the best alignment between a subset of proteins and then gradually extend it by adding more proteins using a greedy strategy. This may impact alignment quality since errors introduced at an earlier stage cannot be fixed later. 

This paper presents a novel one-to-one GNA algorithm, denoted as ConvexAlign, to align multiple PPI networks using a new scoring scheme that integrates network topology, sequence similarity and interaction conservation score. 
It is NP-hard to optimize such a scoring function.
We formulate this GNA problem as an integer program and relax it to a convex optimization problem, which enables us to simultaneously align all the PPI networks, without resorting to the widely-used seed-and-extension or progressive alignment methods. 
Then we use an ADMM (alternating direction method of multipliers) method (see \url{http://stanford.edu/~boyd/admm.html}) to solve the relaxed convex optimization problem and optimize all the protein mappings together. 
Tested on the PPI networks of five different species, ConvexAlign outperforms several popular methods such as IsoRankN, SMETANA,  NetCoffee and BEAMS in terms of biological alignment quality. ConvexAlign finds a few conserved complexes among these 5 species which cannot be found by the other methods.
ConvexAlign also performs very well in aligning some publicly available synthetic networks.
 
\section{Method}
\noindent
\textbf{Definition.} We represent a protein-protein interaction network by an undirected graph $G=(V,E)$ where $V$ is the set of vertices (proteins) and $E$ the set of edges (interactions). Let $d(u)$ denote the degree of vertex $u$ and $e=(u,v)\in E$ represent an edge. A one-to-one global alignment $\set{A}$ between $N$ networks $G_i = (V_i, E_i), 1\leq i \leq N$, is given by a decomposition of all nodes $\mathcal{V} = \cup_{i=1}^{N} V_i$ such that $\mathcal{V} = \set{A}_1\cup \cdots \cup \set{A}_{K}$ where each $\set{A}_i$ contains at most one protein from each network and any two $\set{A}_i$ and $\set{A}_j$ are disjoint. We call each $A_i$ in the alignment a group or a cluster. Proteins in each cluster are mutually aligned to one another.

\subsection{Scoring function for network alignment}
Our goal is to find an alignment that maximizes the number of preserved edges and the number of \textit{matched} orthologous (or functionally conserved) proteins. For this purpose we use a node score for scoring \textit{matched} proteins and an edge score for scoring \textit{matched} interactions, respectively. 
For a pair of proteins, their node score is the combination of their topology score and sequence similarity score. 
We use a minimum-degree heuristic algorithm to calculate the topological score, which was used by us to develop a pairwise GNA method HubAlign~\cite{hashemifar2014hubalign}. A recent third-party evaluation by Pr{\u z}ulj group~\cite{malod2015graal} has shown that this topological score works very well in pairwise GNA. Please see our paper~\cite{hashemifar2014hubalign} for more details. We use the normalized BLAST bit scores for sequence similarity. Let $B(v_i,v_j)$ and $T(v_i,v_j)$ respectively denote the sequence similarity and topology score between a pair of proteins $v_i \in V_i$ and $v_j \in V_j$. Then the node score  $\textup{node}(v_i, v_j)$ is calculated as follows: 
\begin{equation}
\textup{node}(v_i, v_j) = (1-\lambda_1)\textup{B}(v_i, v_j) + \lambda_1\textup{T}(v_i,v_j),
\end{equation}
where  $\lambda_1$ controls the importance of the topology score relative to the BLAST score.
The node score of multiple alignment $\set{A}$, i.e. $f_{node}(\set{A})$, sums the  scores among all pairs of \textit{matched} proteins:
\begin{equation}
f_{node}(\set{A}) = \sum\limits_{1 \leq i < j \leq N} \sum\limits_{\set{A}_k \in \set{A}, v_i, v_j \in \set{A}_{k}}\textup{node}(v_i, v_j).
\end{equation}
%Such a score reflects both sequence and functional similarities between proteins. 
The edge score $f_{interaction}\set(A)$ measures interaction-preserving in an alignment $\set{A}$. This score counts the number of interactions aligned between all pairs of networks:    
\begin{equation}
f_{interaction}(\set{A}) = \sum\limits_{1 \leq i < j \leq N}\sum\limits_{\set{A}_k, \set{A}_{l} \in \set{A}, v_i, v_j \in \set{A}_{k}, v_i', v_j' \in \set{A}_{l}}\delta ((v_i, v_i')\in E_i)\delta ((v_j, v_j')\in E_j),
\end{equation}
where $\delta ((v_i, v_i')\in E_i)$ is an indicator function.
%
%\begin{equation*}
%\delta ((v_i, v_i')\in E_i) = \left\{\begin{array}{lll}
%             1 &  (v_i, v_i')\in E_i\\
%             0 & otherwise
%            \end{array}\right.
%\end{equation*} 
%This score favors alignment of two interactions, which may help identify conserved pathways/complexes. 
We aim to find the multiple alignment $\set{A}$ that maximizes a combination of node and interaction scores as follows.
\begin{equation}
f = (1-\lambda_2)f_{node}(\set{A}) + \lambda_2 f_{interaction}(\set{A}),
\label{Obj:1}
\end{equation}
where $\lambda_2$ describes the tradeoff. See Appendix for determination of $\lambda_1$ and $\lambda_2$ by cross-validation.

\subsection{Integer and Convex Programming Formulation}

\noindent
\textbf{Definition.} A one-to-one multiple network alignment is valid or feasible if the following condition (also called consistency property) is satisfied: for any three vertices $v_i$, $v_j$, and $v_k$ of three different networks, if $v_i$ is aligned to $v_j$ and $v_j$ aligned to $v_k$, then $v_i$ is aligned to $v_k$.

\para{Parameterizing multiple alignments.} Let $M$ be the number of proteins in all the input PPI networks i.e. $M = \sum\limits_{i=1}^{N}|V_i|$. We may represent a valid multiple alignment $\set{A}$ by a binary matrix $Y = (Y_1;Y_2;\cdots;Y_{N})\in \{0,1\}^{M\times K}$, where each block $Y_i$ encodes the association between $V_i$ and $\set{A}$. Each row of $Y$ corresponds to one vertex and each column to one alignment cluster. That is, $\forall v_i\in V_i, Y_i(v_i, A_j) = 1$ if and only if $v_i$ is in cluster $A_j$. Let $\bs{1}$ be a vector of appropriate size with all elements being 1. Since $Y$ is a one-to-one alignment, it shall satisfy the following constraints:
\begin{itemize}
\item Each row of $Y$ has exactly one non-zero entry, i.e., $Y\bs{1} = \bs{1}$ .
\item Each column of $Y$ has at most $N$ non-zero entries, i.e., $Y^{T}\bs{1}\leq N\bs{1}$.
\item Each column of $Y_i$ has at most one non-zero entry, i.e., $Y_i^{T}\bs{1}\leq \bs{1}$.
\end{itemize}
On the other direction, any binary matrix $Y$ satisfying the above properties encodes a one-to-one alignment.

Although $Y$ is a good representation of an MNA, the objective function with $Y$ as variable is nonlinear and thus, hard to optimize. 
Inspired by~\cite{Huang:2013:CSM}, we introduce another alignment matrix $X$ as follows.
\begin{equation}
X = \left(
\begin{array}{cccc}
I_{|V_1|} & X_{12} & \cdots & X_{1N} \\
X_{12}^{T} & I_{|V_2|} & \cdots & X_{2N} \\
\vdots & \cdots & \ddots & \vdots \\
X_{1N}^{T} & \cdots & \cdots & I_{|V_{N}|}
\end{array}
\right) = \left(
\begin{array}{c}
Y_1 \\
Y_2 \\
\vdots \\
Y_{N}
\end{array}
\right)\cdot
\left(
\begin{array}{cccc}
Y_1^{T} & Y_2^{T} & \cdots & Y_{N}^{T}
\end{array}
\right),
\end{equation}
where each block $X_{ij} = Y_i Y_{j}^{T}$ is a binary matrix encoding the mapping between $V_i$ and $V_j$. That is, $X_{i,j}(v_i,v_j)=1$ if and only if $v_i$ and $v_j$ are aligned (i.e., in the same alignment cluster). 

It is easy to see that $X$ is positive semi-definite. Since this section considers only one-to-one mapping, for any two $i$ and $j$ $(i\neq j)$, each row or column of $X_{ij}$ has at most one non-zero element, i.e., $X_{ij}\bs{1}\le \bs{1}$ and $X_{ij}^T\bs{1}\le \bs{1}$ where $\bs{1}$ is a vector of appropriate size with all entries 1. On the other direction, we have the following proposition (see Appendix \ref{proof1} for its proof).

\begin{prop}
Let $X$ be a binary block matrix with $N\times N$ blocks, and $X_{ij}$ be the block in the $i^{th}$ row and the $j^{th}$ column. If $X$ satisfies the following conditions: (1) $X \succeq 0$, (2)$X_{ii} = I_{|V_i|}$ for $1\leq i \leq N$, and (3)$X_{ij}\bs{1} \leq \bs{1}$ and $X_{ij}^{T}\bs{1} \leq \bs{1}$ for  $1\leq i < j \leq N$,
%\begin{align}
%X_{ij}\bs{1} \leq \bs{1}, \ X_{ij}^{T}\bs{1} \leq \bs{1},  & 1\leq i < j \leq N, \nonumber \\
% \nonumber \\
%, \qquad& \nonumber
%\end{align}
then $X$ encodes a feasible global alignment of $N$ networks admitting one-to-one mapping and satisfying the cycle consistency property.
%there exists a matrix $Y$ such that $ X = YY^{T}$.
\label{Prop1}
\end{prop}
Therefore, we may encode a one-to-one GNA using $X$, which leads to a linear formulation of the objective function.
Following Prop.~\ref{Prop1}, we impose the following constraints on $X$:
\begin{align}
X_{ij}\bs{1} \leq \bs{1}, \ X_{ij}^{T}\bs{1} \leq \bs{1}, X_{ij} \in \{0,1\}^{|V_i|\times |V_j|} \qquad (& 1\leq i < j \leq N) \nonumber \\
%X \succeq 0, X_{ii} = I_{|V_i|}, \qquad \qquad & 1\leq i \leq N \nonumber \\
X \succeq 0, \qquad X_{ii} = I_{|V_i|} \qquad(& 1\leq i \leq N) 
%X \succeq 0. \qquad \qquad&
\label{Cons:0}
\end{align}
\para{Objective function.} As $X_{ij}$ is the indicator submatrix for $V_i$ and $V_j$, the node score can be formulated as follows.
\begin{align}
%f_{node} &= \sum\limits_{1\leq i < j \leq N}\sum\limits_{\set{A}_k \in \set{A}, v_i, v_j \in \set{A}_{k}}\textup{node}(v_i, v_j)\nonumber \\
% & = \sum\limits_{1\leq i < j \leq N}\sum\limits_{v \in V_i, v' \in V_j}\textup{node}(v, v') X_{ij}(v,v')
%\nonumber \\
%& = \sum\limits_{1\leq i < j \leq N}\langle C_{ij}, X_{ij} \rangle,
f_{node} = \sum\limits_{1\leq i < j \leq N}\sum\limits_{v \in V_i, v' \in V_j}\textup{node}(v, v') X_{ij}(v,v')
 = \sum\limits_{1\leq i < j \leq N}\langle C_{ij}, X_{ij} \rangle,
\label{Obj:2}
\end{align}
where $C_{ij}$ is a matrix composed of the values of $\textup{node}(v, v')$.

To formulate $f_{interaction}$, we introduce indicator variables $y_{ij}(v_i, v_j, v_i', v_j')$ for edge correspondences:
\begin{equation}
y_{ij}(v_i, v_j, v_i', v_j') = X_{ij}(v_i, v_j) X_{ij}(v_i', v_j'), \quad \forall (v_i, v_i')\in E_i, (v_j, v_j')\in E_j, 1\leq i < j \leq N.
\label{Cons:1}
\end{equation}
\begin{align}
%f_{interaction} & = \sum\limits_{1\leq i < j \leq N}\sum\limits_{(v_i,v_i')\in E_i, (v_j, v_j')\in E_j} y_{ij}(v_i, v_j, v_i', v_j') \nonumber \\
%& = \sum\limits_{1\leq i < j \leq N}\sum\limits_{(v_i,v_i')\in E_i, (v_j, v_j')\in E_j} \langle \bs{1}, \bs{y}_{ij} \rangle,
f_{interaction} = \sum\limits_{1\leq i < j \leq N}\sum\limits_{(v_i,v_i')\in E_i, (v_j, v_j')\in E_j} y_{ij}(v_i, v_j, v_i', v_j') 
 = \sum\limits_{1\leq i < j \leq N} \langle \bs{1}, \bs{y}_{ij} \rangle,
\label{Obj:3}
\end{align}
where $\bs{y}_{ij}$ stacks the indicator variables  between $V_i$ and $V_j$.

The nonlinear constraint~(\ref{Cons:1}) can be replaced by the following linear inequalities (c.f.~\cite{Kumar08ananalysis,Huang:2011:JSS}):
\begin{align}
\forall v_j'\in V_j, \sum\limits_{v_i': (v_i, v_i') \in E_i} y(v_i, v_j, v_i', v_j') \leq X_{ij}(v_i, v_j) \nonumber \\
\forall v_i'\in V_i, \sum\limits_{v_j': (v_j, v_j') \in E_j} y(v_i, v_j, v_i', v_j') \leq X_{ij}(v_i, v_j) \nonumber \\
\forall v_j \in V_j, \sum\limits_{v_i: (v_i, v_i') \in E_i} y(v_i, v_j, v_i', v_j') \leq X_{ij}(v_i', v_j') \nonumber \\
\forall v_i\in V_i, \sum\limits_{v_j: (v_j, v_j') \in E_j} y(v_i, v_j, v_i', v_j') \leq X_{ij}(v_i', v_j')
\label{Cons:2}
\end{align}
It is easy to prove that (\ref{Cons:1}) implies (\ref{Cons:2}).
On the other direction, considering that the coefficients of $\bs{y}$ is positive and we want to maximize (\ref{Obj:3}), we shall be able to prove that (\ref{Cons:2}) implies (\ref{Cons:1}). We replace (\ref{Cons:1}) by (\ref{Cons:2}) to obtain linear constraints and  summarize (\ref{Cons:2}) in the matrix form as follows.
\begin{equation}
B_{ij}\bs{y}_{ij} \leq \set{F}_{ij}(X_{ij}),
\label{Cons:3}
\end{equation}
where $B_{ij}$ is coefficient and $\set{F}_{ij}$ is a linear operator that picks the corresponding element of $X_{ij}$ for each constraint. That is, $\set{F}_{ij}(X_{ij}(v_i,v_j))=<P_{ij},X_{ij}>$ where $P_{ij}$ is a binary matrix with the same dimension as $X_{ij}$ and only one element $P_{ij}(v_i,v_j)$ is equal to 1.

Finally, by integrating (\ref{Obj:2}), (\ref{Obj:3}), (\ref{Cons:3}) and Prop.~\ref{Prop1}%, (\ref{Obj:1})
, we have the following integer program:
\begin{align}
\textup{maximize} & \quad  \sum\limits_{1\leq i < j \leq N}\Big((1-\lambda_2)\langle C_{ij}, X_{ij}\rangle + \lambda_2 \langle \bs{1}, \textbf{y}_{ij}\rangle\Big) & \nonumber \\
\textup{subject to} & \quad \bs{y}_{ij} \in \{0,1\}^{|E_i|\times |E_j|}, \ B_{ij} \bs{y}_{ij} \leq \set{F}_{ij}(X_{ij}), & 1\leq i < j \leq N \nonumber \\
& \quad X_{ij}\bs{1} \leq \bs{1}, \ X_{ij}^{T}\bs{1} \leq \bs{1}, X_{ij} \in \{0,1\}^{|V_i|\times |V_j|}, & 1\leq i < j \leq N\nonumber \\
& \quad X \succeq 0, \quad X_{ii} = I_{|V_i|}, \quad 1\leq i \leq N 
%& \quad X \succeq 0.
\label{MA:IPR}
\end{align}
%If we relax $X$ and $\bs{y}$ to be real-valued between 0 and 1, then the above problem becomes a convex optimization problem. 
The key constraint  is  $X \succeq 0$, which enforces cycle consistency in the alignments. $X \succeq 0$ still holds even $N=2$.

\subsection{\large{Optimization via Convex Relaxation}}
\label{Relax:Opt}
It is NP-hard to directly optimizing (\ref{MA:IPR}) since the variables are binary. We may first relax them to obtain a convex optimization problem that can be solved to global optimum within polynomial time, and then employ a greedy rounding scheme  to convert fractional solution to integral. 

\para{Convex relaxation.} By relaxing $\bs{y}_{ij}$ and $X_{ij}$ to real values between $0$ and $1$, we have the following convex program:
\begin{align}
\textup{maximize} & \quad \sum\limits_{1\leq i < j \leq N}\Big((1-\lambda_2)\langle C_{ij}, X_{ij}\rangle + \lambda_2 \langle \textbf{1}, \textbf{y}_{ij}\rangle\Big) &\nonumber \\
\textup{subject to} & \quad \bs{y}_{ij} \geq \bs{0}, \ B_{ij}\bs{y}_{ij}\leq \set{F}_{ij}(X), & 1\leq i < j \leq N \nonumber \\
 & \quad X_{ij}\bs{1} \leq \bs{1}, \ X_{ij}^{T}\bs{1} \leq \bs{1}, \ X_{ij} \geq 0, & 1\leq i < j \leq N \nonumber \\
                    & \quad X \succeq 0, \quad X_{ii} = I_{|V_i|}, & 1\leq i \leq N 
%                    & \quad X \succeq 0.
\label{MA:SDP}
\end{align}

\para{Optimization strategy.}
We use ADMM (alternating direction of multiplier method) to solve the convex relaxation (\ref{MA:SDP}). The basic idea is to augment its Lagrangian dual(see \url{https://en.wikipedia.org/wiki/Augmented_Lagrangian_method}) and iteratively optimize a subset of variables while keeping the others fixed. This  allows us to exploit structure patterns in the constraint set for effective optimization. As the derivation is quite technical, we leave the details in Appendix \ref{ADMM:Opt}.

\para{Rounding into an integer solution.} The above convex relaxation has a pretty tight fractional solution.  We propose a greedy rounding strategy to convert fractional solution to integral. We collect all the protein pairs with an indicator value $X(u,v)> 0.05$ and place them in a decreasing order into a sorted list $\set{X}$. Then we build an alignment graph starting with an empty edge set by scanning through $\set{X}$. For each scanned protein pair $(u,v)$ in $\set{X}$, in the alignment graph we add an edge to connect this pair as long as such an addition does not violate the constraint that no protein in one network is aligned to two proteins in another network. After all pairs are scanned, we decompose the alignment graph into connected components, each corresponding to a cluster of mutually-aligned proteins. The set of all the clusters form an alignment. Most components are cliques. For the very few non-clique components, we just add some edges to make them cliques. 

\section{RESULTS}
We compare our algorithm ConvexAlign with several popular and publicly available methods IsoRankN~\cite{liao2009isorankn}, SMETANA~\cite{sahraeian2013smetana}, NetCoffee~\cite{hu2013netcoffee} and BEAMS~\cite{alkan2014beams}.  We ran SMETANA and NetCofee with their default parameters. For both BEAMS and IsoRankN, we set three different values for their parameter $\alpha=\{0.3, 0.5, 0.7\}$. We left other parameters of BEAMS at their default values. 
\subsection{Test data}
We use the PPI networks of H.sapiens (human), S.cerevisiae (yeast), Drosophila melanogaster (fly), Caenorhabditis elegans (worm) and Mus musculus (mouse) taken from IntAct~\cite{kerrien2011intact}.  The human network has 9003 proteins and 34935 interactions, the yeast network has 5674 proteins and 49830 interactions, the fly networks has 8374 nodes and 25611 interactions, the mouse network has 2897 proteins and 4372 interactions and the worm network has 4305 proteins and 7747 interactions. Only experimentally-validated PPIs are used.

We also use the NAPAbench~\cite{sahraeian2012network} synthetic PPI networks.  NAPAbench is a benchmark that contains PPI network families generated by three different network models: crystal growth(CG)~\cite{kim2008age}, duplication-mutation-complementation(DMC)~\cite{vazquez2003modeling} and duplication with random mutation(DMR)~\cite{sole2002model}. We use the 8-way alignment dataset of this benchmark, which contains three network families each with 8 networks of 1000 nodes generated by one of the three network models. The 8-way alignment dataset simulates a network family of closely-related species, so this benchmark has very different properties as the above 5 real PPI networks. NAPAbench has recently been used to benchmark SMETANA.

\subsection{Alignment quality measures}
We evaluate multiple network alignment quality using several topological and functional consistency metrics proposed in different studies. Functional consistency measures however, are more important than topological measures since one of the important applications of network alignment is to functional annotation transfer. For topological analysis of the output clusters we use the following metrics.\\
\noindent
\textbf{$c$-coverage}: It is the number of clusters composed of proteins from exactly $c$ species. Specifically, total coverage is the number of clusters composed of proteins from at least two species. Clusters with large $c$ explain a larger amount of data better than clusters with small $c$.

\noindent
\textbf{Conserved Interaction(CI)}: It is calculated as the ratio of the number of aligned interactions to the total number of interactions between output clusters. 

A multiple alignment with a higher $c$-coverage (or total coverage) or CI is not necessarily biologically  meaningful since it may align many unrelated proteins together. Therefore, we also employ GO terms to measure functional consistency or biological quality of an alignment. GO terms describe roles of proteins in terms of their associated biological process (BP), molecular function (MF) and cellular component (CC). We exclude root GO terms from analysis, i.e., GO terms on level higher than 5. We also exclude CC because proteins with matched CC are not usually considered functionally similar. Moreover, CC only annotates a small percentage of the proteins. The following measures are based on the observation that functionally related proteins are more likely to have similar GO terms. 

\noindent
\textbf{Specificity}: We say a cluster $annotated$ if at least two of its proteins have GO annotations. An annotated cluster is $consistent$ if all of its proteins share at least one common GO term. Specificity is defined as the ratio of consistent clusters to annotated clusters. 

\noindent
\textbf{Average of functional similarity ($\overline{AFS}$)}. This score is based on the semantic similarity of the GO terms, which is derived from their distance in the ontology. We use Schlicker\textquotesingle s similarity, based on the Resnik ontological similarity, to calculate the functional similarity in the BP and MF category~\cite{schlicker2006new}. Schlicker\textquotesingle s similarity is one of the best performing methods for computing the functional similarity between proteins (see Appendix~\ref{schilcker} for more details)~\cite{pesquita2009semantic}. Let $s_{cat}(u,v)$ denote the GO functional similarity of proteins $u$ and $v$ in category $cat$ (i.e. BP or MF). $AFS$ of an output cluster $\set{A}$ in category $cat$ is defined as follows:
\begin{equation}
AFS_{cat}(\set{A}) = \frac{1}{\frac{|\set{A}|\times(|\set{A}|-1)}{2}}\sum_{v_i,v_j\in\set{A},i\neq j}s_{cat}(v_i,v_j)\nonumber\
\end{equation}    
Finally we define  $\overline{AFS}$ in category $cat$ as the average of $AFS$ over all clusters. 
Following~\cite{hu2013netcoffee}, we take into consideration all the clusters that contain at least $60\%$ GO-annotated proteins to avoid ignoring many functionally meaningful clusters. 
We separately compare the $\overline{AFS}$ for clusters for $c = 3,4,5$.  We also provide the distribution of the AFS scores for each given $c$. 

\noindent
\textbf{Mean normalized entropy ($MNE$)}. The normalized entropy of a cluster $\set{A}$ is defined as: $NE(\set{A})=\frac{1}{\log(d)}\times\sum_{i=1}^dp_i\times\log(p_i)$ where $d$ is the number of different GO annotations in $\set{A}$ and $p_i$ represents the fraction of proteins in $\set{A}$ with annotation $GO_i$.  A cluster with lower entropy is more functionally coherent. $MNE$ is the mean of normalized entropy over all annotated clusters.

\noindent
\textbf{Conserved orthologous interactions(COI)}: Similar to SMETANA, COI  is calculated as the total number of interactions between all consistent clusters. COI may be a better measure than CI because it detects whether the conserved interactions are spurious or actually correspond to real conserved interactions between orthologous proteins. An alignment with larger COI may lead to identifying functionally conserved subnetwroks (i.e clusters) composed of orthologous genes.  

%\textcolor{red}{Based on two papers that are using this measure, it seems GO terms without closest clusters do not contribute in sensitivity}\\
\noindent
\textbf{Sensitivity}: The closest cluster of a given GO term is the cluster that contains the maximum number of proteins associated with that GO term.  Similar to BEAMS~\cite{alkan2014beams}, we define Sensitivity as the average (over all GO terms) of the fraction of proteins in the closest cluster, that are associated with that GO term. 

\subsection{Alignment quality on real data}
\para{Topological quality.} Table~\ref{quan} lists the topological evaluation of the alignments produced by different methods. The first four multi-rows show the results for the clusters consisting of proteins belonging to $c=2,3,4,5$ species, respectively. In each multi-row, the top and bottom rows show $c$-coverage and the number of proteins in the clusters, respectively. ConvexAlign has a larger $c$-coverage when $c={4,5}$ than the other methods except SMETANA and NetCoffee. However, as we show later, many of clusters generated by these two methods are not functionally conserved. The total coverage of BEAMS and IsoRank is better than the others because they produce many clusters composed of proteins from 2 or 3 species. These clusters can not explain the data as well as clusters containing proteins from 4 or 5 species can. ConvexAlign has a better CI than all other methods except SMETANA. These conserved interactions may be very helpful in identifying the functional modules conserved among networks of different species. It is worth mentioning that most of the conserved interaction resulting from SMETANA may be spurious~\cite{alkan2014beams}.
\begin{table}[h]
\caption{Topological evaluation of output clusters by different alignment methods. IsoRankN and BEAMS are tested using three different values of their parameters $\alpha$.}
\vspace{-8pt}
\label{quan}
\begin{center}
\small
\tabcolsep=0.11cm 
\begin{tabular}{|c|c|c|c|c|c|c|c|c|c|c|c|}
\hline
 & IsoRankN & IsoRankN & IsoRankN & SMETANA  & NetCoffee &BEAMS & BEAMS& BEAMS& ConvexAlign \\
 & (0.3) &(0.5) & (0.7) &   & & (0.3) & (0.5) & (0.7) & \\
\hline
c=2 & 4625 & 4178 & 4670 & 1127 & 1424  &5703 & 5274 & 5271& 2856 \\
 & 11035 & 8356 & 11165 & 2718 &  2848 &11406 & 11469 & 11465& 5712\\
\hline
c=3 & 2259 & 2270 & 2304 & 1653 & 1739& 2192 & 2557 & 2556& 1833\\
 & 8521 & 6810 & 8750 & 5808 &  5217&6576 & 8128 & 8118 &5499\\
\hline
c=4 & 1023 & 731 & 944 & 2028 &  1980 &1163 & 1141 & 1143 & 1190\\
 & 5276 & 2924 & 4823 & 9531 &  7920&4652 & 4686 & 4701& 4760\\
\hline
c=5 & 224 & 112 & 184 & 1622 &  1217&683 & 600 & 600 & 765\\
 & 1417 & 560 & 1182 & 10342 & 6075 & 3915 & 3046 & 3044 &3825\\
\hline
Total  & 8131 & 7291 & 8102 & 6430 &  6360&9741 & 9572 & 9570 & 6644\\
coverage & 26249 & 18650 & 25920 & 28399 &  22070&26549 & 27329 & 27328&19796 \\
\hline
%CI & 4081 & 1402 & 4225 & 12542 &  230&3128 & 3205 & 3195&  2812\\
%\# interactions & 125382 & 58669 & 122959 & 115700 & 27490 &101535 & 106062 & 106068 & 67212\\
CI & 0.03 & 0.02 & 0.03 &0.10& 0.03 &  0.03 & 0.03 & 0.03& 0.04\\
\hline
CIQ & 0.03 & 0.02 & 0.02 &0.06& 0.02 &  0.02 & 0.02 & 0.02& 0.03\\
\hline
\end{tabular}
\end{center}
\vspace{-15pt}
\end{table}

\para{Biological quality.} Table~\ref{func} provides the functional consistency measures of the alignments generated by different methods. The first four multi-rows show the quality of the clusters composed of proteins from $c=2,3,4,5$ species. In these multi-rows, the top and middle rows show the number of consistent and annotated clusters, respectively, and the bottom row shows specificity. Regardless of $c$, ConvexAlign outperforms the other methods in terms of specificity and the number of consistent clusters. At the same time, ConvexAlign generates fewer annotated clusters than BEAMS when $c=2,3,4$. , Although SMETANA and NetCoffee generate a larger number of clusters for $c=4,5$ than ConvexAlign, their clusters are not very functionally consistent. The fifth row shows ConvexAlign has much higher specificity than the others when all the resulting clusters ($c\geq2$) are considered. These results suggest that ConvexAlign finds more functionally consistent clusters, not only by generating small clusters (i.e. $c=2,3$) but more importantly large clusters (i.e. $c=4,5$).  These clusters (especially when $c=4,5$) are very valuable because they may provide useful information about the orthology relationship among the proteins of all species. Moreover, these clusters can be very useful for identifying conserved sub-networks as well as predicting the function of unannotated proteins. ConvexAlign yields a $COI/CI$ ratio around $60\%$ that is $1.44$ times larger than the second best ratio by BEAMS. This result may indicate that ConvexAlign is able to identify conserved interactions between othologous proteins. 
It also suggests that although SMETANA has the largest CI, many of those conserved interactions are possibly false and formed by non-orthologous proteins. ConvexAlign also outperforms other methods in terms of $MNE$ and sensitivity.
\begin{table}[!h]
\caption{Functional consistency of output clusters. Note that for MNE, the smaller the better; while for the other measures, the larger the better.}
\label {func}
\vspace{-15pt}
\begin{center}
\small
\tabcolsep=0.11cm 
\begin{tabular}{|c|c|c|c|c|c|c|c|c|c|c|}
\hline
& &IsoRankN & IsoRankN & IsoRankN & SMETANA  & NetCoffee &BEAMS & BEAMS& BEAMS&ConvexAlign\\
 & &(0.3) &(0.5) & (0.7) &   & & (0.3) & (0.5) & (0.7)  \\
\hline
 & consistent & 906 & 1259 & 919 & 295 &  495&1539 & 1568 & 1569 & 1914\\
 c=2& annotated & 3614 & 2862 & 3646 & 931 &  931 &3486 & 3456 & 3452 & 2326\\
 & specifity & 0.25 & 0.44 & 0.25 & 0.39 &  0.53 &0.44 & 0.45 & 0.45&0.82\\
\hline
 & consistent & 203 & 466 & 231 & 188 & 462& 1003 & 1084 & 1084  & 1155\\
c=3 & annotated & 2160 & 2153 & 2210 & 1556 & 1640& 2119 & 2442 & 2441 &1741\\
 & specifity & 0.09 & 0.22 & 0.10 & 0.12 & 0.28& 0.47 & 0.44 & 0.44& 0.66\\
\hline
& consistent & 41 & 106 & 54 & 170 &  406 & 606&624 & 624  &661\\
c=4  & annotated & 1020 & 723 & 942 & 2019 &1640 & 1159 & 1136 & 1138& 1079\\
 & specifity & 0.04 & 0.15 & 0.06 & 0.08 &  0.25&0.52 & 0.55 & 0.55&0.61\\
\hline
& consistent & 14 & 19 & 9 & 183 &  406 & 383 & 359 & 359 & 493\\
c=5 & annotated & 224 & 112 & 184 & 1621 & 1955 &  683& 600 & 600& 763\\
& specifity & 0.06 & 0.17 & 0.05 & 0.11 & 0.21 & 0.56 & 0.60 & 0.60&0.65\\
\hline
$c\geq2$ & specifity & 0.17 & 0.32 & 0.17 & 0.14 & 0.29 &  0.48 & 0.48 & 0.48 &0.71\\
\hline
 \multicolumn{2}{|c|}{COI}&88&188&127&480&553&1237&1311&1305&1668\\
 \multicolumn{2}{|c|}{COI/CI}&0.02&0.13&0.03&0.04&0.21&0.40&0.41&0.41&0.59\\
\hline
 \multicolumn{2}{|c|}{MNE}&2.15&2.19&2.14&2.44&2.39&1.97&1.95&1.95&1.93\\
\hline
 \multicolumn{2}{|c|}{Sensitivity}&0.45 & 0.46 & 0.45 & 0.36 &0.22 &  0.33 & 0.31 & 0.37  &0.51\\
\hline
\end{tabular}
\end{center}
\vspace{-15pt}
\end{table}

Table \ref{func:ave} shows the $\overline{AFS}$ separately for clusters composed of proteins in 3, 4, and 5 species in both categories BP and MF. The $\overline{AFS}$ obtained by ConvexAlign is $6-20\%$ larger than the other methods. These results indicate that on average the clusters generated by ConvexAlign are functionally more consistent. That is, ConvexAlign outperforms the other methods in terms of not only the number of consistent clusters, but also the average GO semantic similarity.
\begin{table}[h]
\caption{$AFS$ comparison between ConvexAlign and the other methods }
\label {func:ave}
\vspace{-15pt}
\begin{center}
\small
\tabcolsep=0.11cm 
\begin{tabular}{|c|c|c|c|c|c|c|c|c|c|c|}
\hline
& &IsoRankN & IsoRankN & IsoRankN & SMETANA  & NetCoffee &BEAMS & BEAMS& BEAMS&ConvexAlign\\
 & &(0.3) &(0.5) & (0.7) &   & & (0.3) & (0.5) & (0.7)  \\
\hline
&c=3&0.83&1.02&0.86&0.74&1.03&1.60&1.63&1.63&\textbf{1.74}\\
$\overline{AFS}_{BP}$&c=4&0.69&0.97&0.72&0.68&0.99&1.63&1.61&1.60&\textbf{1.79}\\
&c=5&0.75&1.01&0.72&0.85&1.16&1.66&1.67&1.67&\textbf{1.71}\\
\hline
&c=3&0.80&0.94&0.80&0.69&0.99&1.40&1.33&1.34&\textbf{1.54}\\
$\overline{AFS}_{MF}$&c=4&0.83&1.02&0.86&0.74&1.03&1.60&1.63&1.63&\textbf{1.74}\\
&c=5&0.86&1.06&0.86&0.94&1.18&1.68&1.68&1.68&\textbf{1.74}\\
\hline
\end{tabular}
\vspace{-15pt}
\end{center}
\end{table}
The distribution of AFS scores for clusters composed of proteins in 3, 4, and 5 species is shown in Fig.~\ref{fig:semsim}, in which the middle line in each box shows the median value. That is, the median AFS obtained by ConvexAlign is larger than the other methods. These results further confirm that ConvexAlign yields clusters with higher functional similarity in both categories MF and BP.

\subsection{Alignment quality on synthetic data}
\begin{figure}[!ht]
\hspace{0 in}
  \includegraphics[width=1\textwidth]{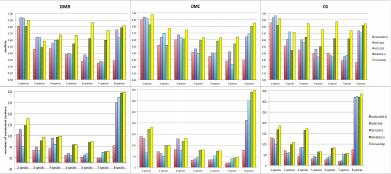} 
\vspace{-15pt}
 \caption{Specificity (top) and the number of consistent clusters (bottom) generated by the competing methods for different $c$ on synthetic data. Only the best performance for IsoRankN and BEAMS is shown.}
\label{fig:spesynth}
\end{figure}

\begin{figure}[!ht]
\hspace{0 in}
\center
  \includegraphics[width=0.7\textwidth]{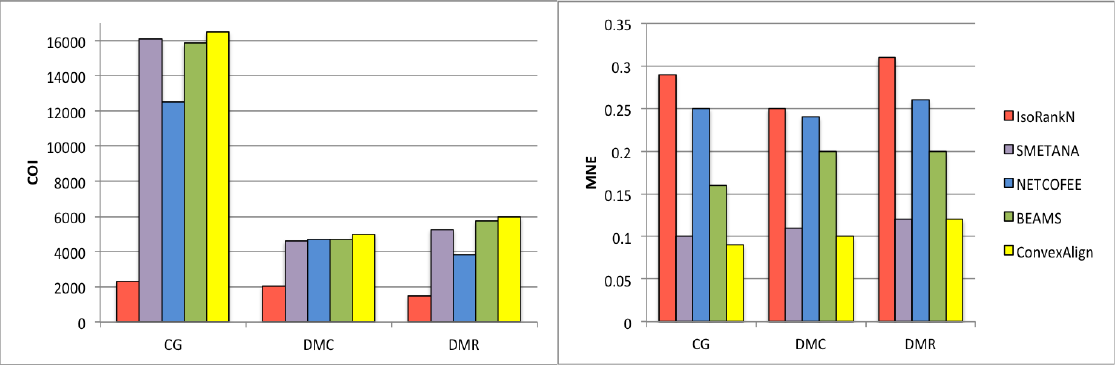} 
\vspace{-10pt}
 \caption{COI and MNE of the clusters generated by the competing methods on synthetic data. Only the best performance for IsoRankN and BEAMS is shown.}
\label{fig:mnesynth}
\end{figure}

This section explains the results on the NAPAbench benchmark. 
Fig.~\ref{fig:spesynth} shows the number of consistent clusters generated by different methods and their specificity on clusters composed of proteins in $c=2,3,4,5,6,7,8$ species, respectively. In terms of the number of consistent clusters, ConvexAlign is slightly better than the second best method BEAMS regardless of $c$, but much better than the others. In terms of specificity, ConvexAlign has a much larger advantage over the other methods when $c=4,5,6,7$. When $c=8$, ConveAlign is slightly better than BEAMS, but much better than the others. These results indicate that ConvexAlign aligns proteins in a functionally  consistent way, without generating too many spurious clusters in which the proteins appear to be unrelated. 
%In a more detailed analysis, Fig.~\ref{fig:spesynth} suggests that ConvexAlign is not only capable of finding functionally consistent clusters with small sizes, but also clusters composed of orthologous proteins in most of the species. 
Fig.~\ref{fig:mnesynth} shows that ConvexAlign outperforms all the other methods in terms of both $MNE$ and $COI$. 
%Due to space limit, the topological evaluation is presented in Figure~\ref{fig:toposynth} and Figure~\ref{fig:cisynth}.
Due to space limit, the topological evaluation is presented in Fig.~\ref{fig:cisynth}.
 
\subsection{Finding conserved subnetworks}
\begin{figure}[!ht]
\hspace{0 in}
  \includegraphics[width=1\textwidth]{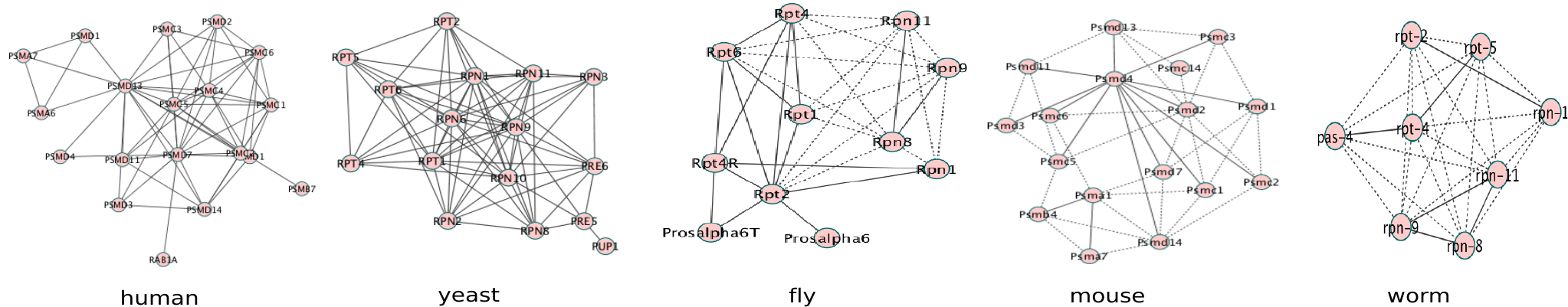} 
 \vspace{-15pt}
 \caption{The ConvexAlign-detected proteasome complex in each input PPI network.}
\label{fig:bio1}
\end{figure}

\begin{figure}[!ht]
\center
\hspace{0 in}
  \includegraphics[width=0.5\textwidth]{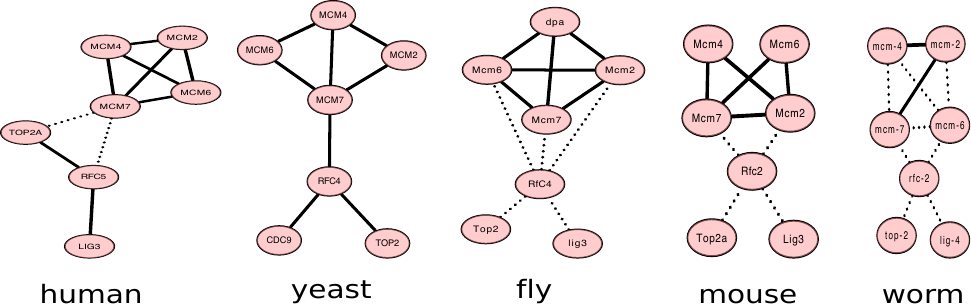} 
  \caption{The ConvexAlign-detected DNA replication complex in each input PPI network.}
\label{fig:bio2}
 \vspace{-15pt}
\end{figure}

One of the applications of network alignment is to reveal subnetworks conserved across the species. These subnetworks may be helpful for extracting biological information that cannot be inferred from sequence similarity alone. 
Fig.~\ref{fig:bio1} shows one conserved complex detected by ConvexAlign among the 5 species: human, yeast, fly, mouse and worm, but not appearing in the alignments generated by other methods. This complex is enriched for proteasome (with $p$-value$<10^{-7}$ in all species), which is essential for the degradation of most proteins including misfolded or damaged proteins. The aligned nodes are shown in Table \ref{tab:bio}. In Fig.~\ref{fig:bio1}, the interactions in IntAct are displayed in solid lines. For fly, mouse and worm, some edges (shown by dotted lines) are missing in IntAct but present in the STRING database~\cite{szklarczyk2011string} with experimental evidence at the highest confidence. Note that our input networks consist of interactions only from IntAct but not STRING. This suggests that ConvexAlign is able to predict missing interactions.
%All the proteins aligned by ConvexAlign are associated with the same GO category, which reflects functional coherence of the predicted clusters. 
We use PANTHER~\cite{mi2010panther} to check if the aligned nodes are orthologous proteins. Most of the aligned proteins are shown to be least divergent orthologs. As shown in Table~\ref{tab:bio}, there are some missing proteins from different species in some of the clusters. This is because either there are no orthologs in those species or there is no alignment for them. For example, cluster $5$ has no proteins from worm and fly. PANTHER could not find any orthologous proteins in those species either. Cluster $6$ misses orthologous proteins in fly and yeast, which are aligned by ConvexAlign to proteins not in this proteasome complex. In addition, this proteasome complex has different number of nodes in different species, which implies that ConvexAlign is able to deal with inserted and deleted nodes. 
Fig.~\ref{fig:bio2} shows another conserved subnetwork detected by ConvexAlign that is related to DNA replication (with $p$-value$<6^{-10}$ in all species). Again, this subnetwork cannot be detected by the other methods. PANTHER suggests that the aligned proteins are orthologous and functionally related (see Table~\ref{tab:bio2}).

\section{Discussion}
This paper has presented a new method ConvexAlign for global alignment of multiple PPI networks. ConvexAlign uses a network alignment scoring function that integrates sequence and topological similarity between the matched proteins as well as interaction consistency. Then ConvexAlign uses a novel convex formulation to simultaneously align all the proteins in multiple input networks, resulting in better alignment quality. Such a formulation allows us to use an ADDM method to find its optimal solution.

We have tested ConvexAlign on both real PPI networks and the synthetic data,  evaluated the output alignments by different performance metrics and compared it to several popular methods. Experimental results on the real data show that on average ConvexAlign generates more functionally consistent clusters consisting of proteins from most of the input species. That is, ConvexAlign can explain a larger amount of data in a more functionally meaningful way. ConvexAlign can also find a few conserved and biologically important complexes which cannot be detected by the other alignment methods.

In the future we may extend ConvexAlign to produce many-to-many global alignments, which will require some revision of our formulation. It will also be interesting to study how to revise our convex formulation for local alignments of multiple PPI networks. Of course we may also apply ConvexAlign to the alignment of other biological systems such as metabolic networks and protein structures.

Currently the time complexity of our algorithm is $O(M^{3}K)$, where $K$ is the number iterations in ADMM and $M$ of the total number of proteins. Using a single computer, it takes dozens of hours to align the real PPI networks of the five species and only 1.5 hours to align 8 synthetic networks.
The $M^3$ factor comes from the eigen-decomposition of a $M\times M$ matrix, incurred by the consistency constraint $X \succeq 0$. We may explore a few strategies to speed up this step. For example, we may place the positive semidefinite constraint on a submatrix of $X$, i.e., enforcing the consistency among only a subset of important nodes (i.e., hub nodes and/or nodes adjacent to hubs). 
since $X$ is sparse and block-structured, we may also apply some block-based or parallel algorithms to speed up eigen-decomposition.

\bibliographystyle{plain}
\bibliography{trans_sync_bib}

\newpage
\appendix{\textbf{\large{APPENDIX}}}
\renewcommand\thefigure{\thesection.\arabic{figure}}  
\renewcommand\thetable{\thesection.\arabic{table}}
\section{Results}
\setcounter{figure}{0}
\setcounter{table}{0}
\label{res:more}

\begin{figure}[!ht]
\hspace{0 in}
  \includegraphics[width=1\textwidth]{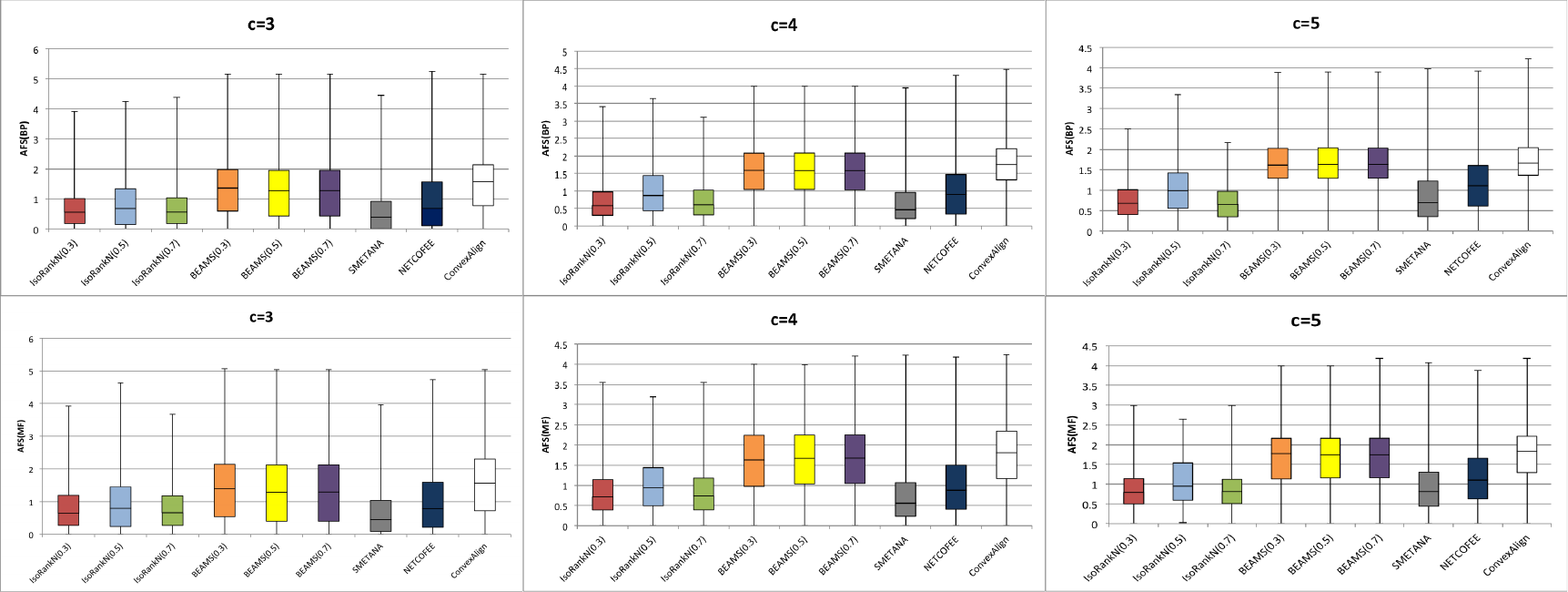} 
 \caption{AFS distribution resulting from different methods on real data.}
\label{fig:semsim}
\end{figure} 

%\begin{figure}[!ht]
%\hspace{0 in}
%  \includegraphics[width=1\textwidth]{topo-synth.pdf} 
% \caption{Coverage analysis on synthetic data. }
%\label{fig:toposynth}
%\end{figure} 

\begin{figure}[!ht]
\center
\hspace{0 in}
  \includegraphics[width=0.5\textwidth]{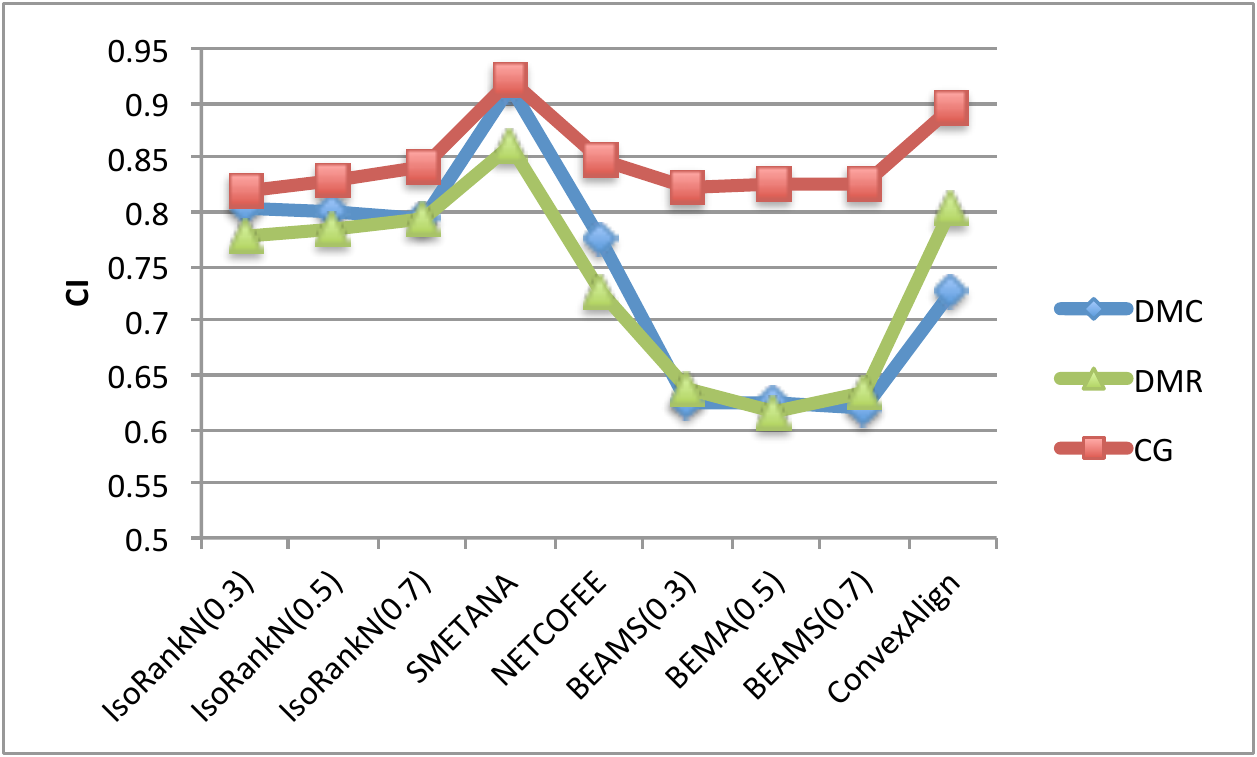} 
 \caption{CI analysis on synthetic data. }
\label{fig:cisynth}
\end{figure} 

\begin{table}[!ht]
\caption{Clusters of aligned nodes that are enriched for proteasome.}
\begin{center}
\small
\tabcolsep=0.11cm 
\begin{tabular}{|c|c|c|c|c|c|c|c|c|c|}
\hline
 & mouse & worm & yeast & fly & human  \\
\hline
cluster 1 & Psmb4 & - & PUP1 & - & PSMB7  \\
\hline
cluster 2 & Psmd7 & rpn-8 & RPN8 & Rpn8 & PSMD7  \\
\hline
cluster 3 & Psma1 & - & PRE5 & Prosalpha6T & PSMA6  \\
\hline
cluster 4 & Psmd14 & rpn-11 & RPN11 & Rpn11 & PSMD14  \\
\hline
cluster 5 & Psmd1 & - & RPN2 & - & PSMD1  \\
\hline
cluster 6 & Rab1A & - & - & - & RAB1A  \\
\hline
cluster 7 & Psmc6 & rpt-4 & RPT4 & Rpt4 & PSMC6  \\
\hline
cluster 8 & Psmc3 & rpt-5 & RPT5 & Rpt4R & PSMC3  \\
\hline
cluster 9 & Psmd2 & rpn-1 & RPN1 & Rpn1 & PSMD2  \\
\hline
cluster 10 & Psma7 & pas-4 & PRE6 & Prosalpha6 & PSMA7  \\
\hline
cluster 11 & Psmc2 & - & RPT1 & Rpt1 & PSMC2  \\
\hline
cluster 12 & Psmc1 & rpt-2 & RPT2 & Rpt2 & PSMC1  \\
\hline
cluster 13 & Psmc5 & - & RPT6 & Rpt6 & PSMC5  \\
\hline
cluster 14 & Psmd13 & rpn-9 & RPN9 & Rpn9 & PSMD13  \\
\hline
cluster 15 & Psmd11 & - & RPN6 & - & PSMD11  \\
\hline
cluster 16 & Psmd3 & - & RPN3 & - & PSMD3  \\
\hline
cluster 17 & Psmd4 & - & RPN10 & - & PSMD4  \\
\hline
\hline
\end{tabular}
\end{center}
\vspace{-15pt}
\label{tab:bio}
\end{table}

\begin{table}[!ht]
\caption{Clusters of aligned nodes that are enriched for DNA replication.}
\begin{center}
\small
\tabcolsep=0.11cm 
\begin{tabular}{|c|c|c|c|c|c|c|c|c|c|}
\hline
 & mouse & worm & yeast & fly & human  \\
 \hline
cluster 1&Top2a&top-2&CDC25C&Top2&TOP2A\\
\hline
cluster 2&Mcm2&mcm-2&MCM2&Mcm2&MCM2\\
\hline
cluster 3&Mcm4&mcm-4&MCM4&dpa&MCM4\\
\hline
cluster 4&Mcm6&mcm-6&MCM6&Mcm6&MCM6\\
\hline
cluster 5&Mcm7&mcm-7&MCM7&Mcm7&MCM7\\
\hline
cluster 6&Lig3&lig-4&CDC9&lig3&LIG3\\
\hline
cluster 7&Rfc2&rfc-2&RFC4&RfC4&RFC5\\
\hline
\hline
\end{tabular}
\end{center}
\vspace{-15pt}
\label{tab:bio2}
\end{table}

\section{\large{Schlicker\textquotesingle s similarity}}
\label{schilcker}
Given two GO terms $g_1$, $g_2$ and their least common ancestor $g_c$, the Resnik ontological similarity is defined as $sim_Resnik(g_1,g_2 )=IC(g_c)$, where $IC(g)$ is the information content of the term $g$ in the given annotation dataset. Considering two gene products $p$ and $q$ annotated with the sets $GO^p$ and $GO^q$ of GO terms, respectively with sizes $N$ and $M$, a similarity matrix $S$ is calculated such as:
\begin{equation}
s_{ij}=sim_{Resnik}(GO_i^p,GO_j^q ),\forall  i\in\{1,\cdots,N\}, \forall  j\in\{1,\cdots,M\}\nonumber\
\end{equation}
This matrix contains all pairwise similarity values between all GO terms associated to $p$ and $q$. The average over the row maxima and the column maxima, respectively, gives similarity values for the comparison of $p$ to $q$ and the comparison of $q$ to $p$:
\begin{equation}
score_{row}=\frac{1}{N}\sum_{(i=1)}^N max(s_{ij} )_{(1\leq j\leq M)},\\\nonumber
score_{column}=\frac{1}{M}\sum_{(j=1)}^M max(s_{ij} )_{(1\leq i\leq N)}.\nonumber\
\end{equation}
The Schlicker\textquotesingle s similarity is then calculated as $sim_{func }(p,q)=max\{score_{row},score_{column}\}$.

\section{\large{Parameter selection}}
For all the experiments in this paper, we set the parameters as $\lambda_1 = 0.3$, and $\lambda_2 = 0.02$. These parameters are chosen via 10-fold cross-validation in optimizing the GO-term scores of the alignment between the mouse and worm networks. The weight factor for aligned interactions is small because: 1) there are many more aligned interactions than aligned nodes, so a small $\lambda_2$ may place the node and interaction scores at the similar scale; and 2) the topological score used in our scoring function already encodes some interaction information and thus, may overlap with the interaction score. Of course we may increase $\lambda_2$ to favor other performance metrics such as the number of aligned interactions and the number of annotated clusters.

\section{\large{Proof of Proposition~\ref{Prop1}}}
\label{proof1}
\begin{proof}
Let $V_i$ be a vertex set corresponding to all rows in the block $X_{ii}$. Then $X_{ii}$ has size $|V_i|\times V_i$. Each block $X_{ij}$ has size $|V_i|\times |V_j|$, describing the relationship between $V_i$ and $V_j$. Let $V=\cup_{1\leq i \leq N} V_i$, then the binary block matrix $X$ has size $|V|\times |V|$. Ignoring the $N$ blocks $X_{ii}$ ($1\le i\le N$), the binary block matrix $X$ can be treated as an adjacency matrix of a simple graph $\set{M}$.
That is, starting from $X$, we may construct a simple graph $\set{M}$ for $V$ such that one non-zero entry in $X$ corresponds to one edge in $\set{M}$.

According to the constraints $X_{ij}\bs{1}\leq \bs{1}, X_{ij}^{T}\bs{1}\leq \bs{1}$, it is easy to see that each connected component of $\set{M}$ contains at most one vertex from each $V_i$.
Now we want to prove that each connected component in $\set{M}$ is a clique.
This is equivalent to proving the consistency property, i.e., given three vertices $v_i \in V_i, v_j \in V_j, v_k \in V_k$, if $X_{ij}(v_i, v_j) = X_{ik}(v_i, v_k) = 1$, then $X_{jk}(v_j, v_k) = 1$.
This can be induced from that $X$ is positive semidefinite.
Consider the $3\times 3$ principal submatrix of $X$ induced by $v_i, v_j, v_k$. Since the principle submatrix of a positive semidefinite matrix is also semidefinite positive, we have
$$
\left(
\begin{array}{ccc}
1 & 1 & 1 \\
1 & 1 & X_{jk}(v_j,v_k) \\
1 & X_{jk}(v_j, v_k) & 1
\end{array}
\right) \succeq 0.
$$
It follows that
$$
det (\left(
\begin{array}{ccc}
1 & 1 & 1 \\
1 & 1 & X_{jk}(v_j,v_k) \\
1 & X_{jk}(v_j, v_k) & 1
\end{array}
\right)) = - (1- X_{jk}(v_j, v_k))^2.
$$
This implies $X_{jk}(v_j, v_k) = 1$.

Let $\set{A}=\{A_1,A_2,...,A_K\}$ denote all the connected components of $\set{M}$. Then $\set{A}$ is a feasible one-to-one alignment between the $N$ vertex sets $V_1$,$V_2$,...,$V_N$.
For each vertex $V_i$, we construct a binary matrix $Y_i$ of size $|V_i|\times K$. For any $k$ ($1\le k\le K$) and $v_i\in V_i$, if $v_i$ appears in $A_k$, then $Y_i(v_i,k)=1$; otherwise $Y_i(v_i,k)=0$. Finally we construct a $|V|\times K$ binary matrix $Y$ by stacking $Y_1$, $Y_2$,...,$Y_N$ along the vertical direction. It is easy to show that $X=YY^T$.
\end{proof}

\section{\large{Optimization via ADMM (Alternating Direction Method of Multipliers)}}
\label{ADMM:Opt}

In this section, we describe in detail how to solve the optimization problem in (\ref{MA:SDP}) using ADMM. The basic idea is to augment its Lagrangian dual (see ??) and iteratively optimize a subset of variables while keeping the others fixed. This allows us to exploit structure patterns in the constraint set for effective optimization.
%alternatively optimize its augmented Lagrangian, so that in each suboptimization step we only solve a small-scale optimization problem. 
To maximize the power of ADMM, we introduce a latent variable $\bs{s}_{ij}$ to  break each constraint $B_{ij}\bs{y}_{ij} \leq  \set{F}_{ij}(X_{ij})$ into two sets of constraints $B_{ij}\bs{y}_{ij} \leq \bs{s}_{ij}$ and $\bs{s}_{ij} = \set{F}_{ij}(X_{ij})$. Let $\lambda=\frac{\lambda_2}{1-\lambda_2}$ and $\bs{d}_{ij}$ be the coefficient of $\bs{y}_{ij}$. The relaxed convex optimization problem can be written as follows.
\begin{align}
\textup{maximize} & \quad \sum\limits_{1\leq i < j \leq N}\Big(\langle C_{ij}, X_{ij}\rangle + \lambda \langle \textbf{d}_{ij}, \textbf{y}_{ij}\rangle\Big) &\nonumber \\
\textup{subject to} & \quad \bs{y}_{ij} \geq \bs{0}, \ 1\leq i \leq j \leq N &\quad: \bs{z}_{ij}^{0} \geq 0 \nonumber \\
                    & \quad B_{ij}\bs{y}_{ij} \leq \bs{s}_{ij},\ 1\leq i \leq j \leq N  & \quad: \bs{z}_{ij,1} \geq 0\nonumber \\
                    & \quad \bs{s}_{ij} = \set{F}_{ij}(X_{ij}),\ 1\leq i < j \leq N & \quad :\bs{z}_{ij,2} \nonumber \\
                    & \quad X_{ij}\bs{1} \leq \bs{1},\ 1\leq i \leq j \leq N &: \bs{z}_{ij,3} \geq 0 \nonumber \\
                    & \quad X_{ij}^{T}\bs{1} \leq \bs{1},\ 1\leq i \leq j \leq N &: \bs{z}_{ij,4}\geq 0 \nonumber \\
                    & \quad X_{ij} \geq 0,\ 1\leq i \leq j \leq N & : Z_{ij,5} \geq 0\nonumber \\
                    & \quad X_{ii} = I_{|V_i|}, \ 1\leq i \leq N &: Z_{ii,6} \nonumber \\
                    & \quad X \succeq 0 & : S \succeq 0
\end{align}
Note that the right column shows the dual variables of the corresponding constraints.
Using the dual variables, the Lagrangian of the above problem is as follows:
\begin{align}
\set{L} &=  \sum\limits_{1\leq i < j \leq N} \Big(-\langle C_{ij}, X_{ij}\rangle - \lambda  \langle \bs{d}_{ij}, \bs{y}_{ij}\rangle - \langle \bs{z}_{ij,0}, \bs{y}_{ij} \rangle + \langle \bs{z}_{ij,1}, B_{ij} \bs{y}_{ij} - \bs{s}_{ij} \rangle + \langle \bs{z}_{ij,2}, \bs{s}_{ij} - \mathcal{F}_{ij}(X_{ij})\rangle  \nonumber \\
        &  \qquad\qquad + \langle \bs{z}_{ij,3}, X\bs{1} + \bs{1} \rangle + \langle \bs{z}_{ij,4}, X^{T}\bs{1} - \bs{1} \rangle - \langle Z_{ij,5}, X_{ij} \rangle - 2\langle S_{ij}, X_{ij} \rangle \Big) \nonumber \\
        & \quad + \sum\limits_{1\leq i \leq N}\Big(\langle Z_{ii,6}, X_{ii} - I_{|V_i|} \rangle - \langle S_{ii}, X_{ii} \rangle  \Big) \nonumber \\
     & = \sum\limits_{1\leq i < j \leq N} \Big(- \langle \bs{1}, \bs{z}_{ij,3}\rangle - \langle \bs{1}, \bs{z}_{ij,4} \rangle - \langle C_{ij} + \mathcal{F}_{ij}^{T}(\bs{z}_{ij,2})- \bs{z}_{ij,3}\bs{1}^{T} - \bs{1}\bs{z}_{ij,4}^{T} + Z_{ij,5}, X_{ij}\rangle \nonumber \\
        & \quad\quad - \langle \lambda \bs{d}_{ij} + \bs{z}_{ij,0} - B_{ij}^{T}\bs{z}_{ij,1}, \bs{y}_{ij}\rangle  - \langle \bs{z}_{ij,1} - \bs{z}_{ij,2}, \bs{s}_{ij} \rangle \Big) \nonumber \\
        & \quad + \sum\limits_{1\leq i \leq N}\Big(-\langle I_{|V_i|}, Z_{ii,6} \rangle - \langle S_{ii} - Z_{ii,6} ,X_{ii}\rangle \Big). \nonumber \\
\end{align}
The augmented Lagrangian dual of the above problem is as follows.
\begin{align}
\set{L}' & = \sum\limits_{1\leq i < j \leq N} \Big(\langle \bs{1}, \bs{z}_{ij,3}\rangle - \langle \bs{1}, \bs{z}_{ij,4} \rangle + \langle C_{ij} + \mathcal{F}_{ij}^{T}(\bs{z}_{ij,2})- \bs{z}_{ij,3}\bs{1}^{T} - \bs{1}\bs{z}_{ij,4}^{T} + Z_{ij,5} + 2S_{ij}, X_{ij}\rangle \nonumber \\
        & \quad\quad + \langle \lambda \bs{d}_{ij} + \bs{z}_{ij,0} - B_{ij}^{T}\bs{z}_{ij,1}, \bs{y}_{ij}\rangle  + \langle \bs{z}_{ij,1} - \bs{z}_{ij,2}, \bs{s}_{ij} \rangle \nonumber \\
        & \quad\quad +\frac{1}{2\mu}\big(\|C_{ij} + \mathcal{F}_{ij}^{T}(\bs{z}_{ij,2})- \bs{z}_{ij,3}\bs{1}^{T} - \bs{1}\bs{z}_{ij,4}^{T} + Z_{ij,5}+2S_{ij}\|_{F}^2+ \|\lambda \bs{d}_{ij} + \bs{z}_{ij,0} - B_{ij}^{T}\bs{z}_{ij,1}\|^2+\|\bs{z}_{ij,1} - \bs{z}_{ij,2}\|^2\big)\Big) \nonumber \\
        & \quad + \sum\limits_{1\leq i \leq N}\Big(\langle I_{|V_i|}, Z_{ii,6} \rangle +\langle S_{ii} - Z_{ii,6} ,X_{ii}\rangle + \frac{1}{2\mu}\|S_{ii} - Z_{ii,6}\|_{F}^2 \Big).
\end{align}

%\textcolor{red}{Please fix the below sentences. Marked by Jinbo Xu}

$\set{L}'$ shall be maximized with respect to the primal variables but minimized with respect to the dual variables. We initialize all the primal and dual variables to zero. At iteration $k+1$, we update the dual and primal variables as follows.

\para{Step 1: Optimizing $\bs{z}_{ij,0}^{(k+1)}$.} When variable $\bs{z}_{ij,0}$ is active with other variables fixed, the optimization problem is equivalent to computing
$$
\min_{\bs{z}_{ij,0}\geq 0}\|\lambda \bs{d}_{ij} + \bs{z}_{ij,0} - B_{ij}^{T}\bs{z}_{ij,1}^{(k)} + \mu \bs{y}_{ij}^{(k)}\|^2, \quad 1\leq i < j \leq N \nonumber \\
$$
In this case, we have
$$
\bs{z}_{ij,0}^{(k+1)} = \max(0, B_{ij}^{T}\bs{z}_{ij,1}^{(k)} - \mu \bs{y}_{ij}^{(k)} -\lambda \bs{d}_{ij}), \quad 1\leq i < j \leq N
$$

\para{Step 2: Optimizing $\bs{z}_{ij,1}^{(k+1)}$.}  When $\bs{z}_{ij,1}$ is active while other variables are fixed, each of them can be optimized independently as
$$
\min_{\bs{z}_{ij,1}\geq 0}\|B_{ij}^{T}\bs{z}_{ij,1} - \Big(\lambda \bs{d}_{ij} + \bs{z}_{ij,0}^{(k+1)} + \mu \bs{y}_{ij}^{(k)}\Big)\|^2 + \|\bs{z}_{ij,1} - \bs{z}_{ij,2}^{(k)}\|^2.
$$

Since each $\bs{y}$ appears in 4 constraints of (\ref{Cons:2}), we permute the rows of $B_{ij}$ to form 4 submatrices $B_{ij,1}, B_{ij,2}, B_{ij,3}, B_{ij,4}$, so that each column of one submatrix contains exactly one non-zero entry. Accordingly, we can also reorder and rewrite $\bs{z}_{ij,1}$ as $(\bs{z}_{ij,1,1};\bs{z}_{ij,1,2};\bs{z}_{ij,1,3};\bs{z}_{ij,1,4})$. Then we alternate the optimization of $\bs{z}_{ij,1,l}(l=1,2,3,4)$ as follows.

%For efficient optimization, we utilize the division $B_{ij} = \big(B_{ij,1}, B_{ij,2}\big)$ so that each row of $B_{ij,1}$ and $B_{ij,2}$ has exactly one non-zero entry. Introduce the corresponding decomposition of $\bs{z}_{ij,1} = (\bs{z}_{ij,1,1};\bs{z}_{ij,1,2})$. We start from $(\bs{z}_{ij,1,1}^{\star}; \bs{z}_{ij,1,2}^{\star}) = \bs{z}_{ij,1}^{(k)}$ and alternate the optimization of $\bs{z}_{ij,1,1}$ and $\bs{z}_{ij,1,2}$ as follows
\begin{align}
\bs{z}_{ij,1,l}^{(k+1)} = \arg\min_{\bs{z}_{ij,1,l}\geq 0}\|B_{ij,l}^{T}\bs{z}_{ij,1,l} - \Big(\lambda \bs{d}_{ij} + \bs{z}_{ij,0}^{(k+1)} + \mu \bs{y}_{ij}^{(k)}-\sum_{p\neq l}B_{ij,p}^{T}\bs{z}_{ij,1,p}^{\star}\Big)\|^2 + \|\bs{z}_{ij,1,l} - \bs{z}_{ij,2,l}^{(k)}\|^2.
%\bs{z}_{ij,1,2}^{\star} = \arg\min_{\bs{z}_{ij,1,2}\geq 0}\|B_{ij,2}^{T}\bs{z}_{ij,1,2} - \Big(\lambda \bs{d}_{ij} + \bs{z}_{ij,0}^{(k+1)} + \mu \bs{y}_{ij}^{(k)}-B_{ij,1}^{T}\bs{z}_{ij,1,1}^{\star}\Big)\|^2 + \|\bs{z}_{ij,1,2} - \bs{z}_{ij,2,2}^{(k)}\|^2.
\end{align}
where $\bs{z}_{ij,1,p}^{\star}$ is the latest value of $\bs{z}_{ij,1,p}$.
Due to the special structure of $B_{ij,l}(l=1,2,3,4)$, the elements of $\bs{z}_{ij,1,l}$  can be optimized independently, leading to explicit expressions of optimal values:
\begin{align}
\bs{z}_{ij,1,l}^{(k+1)} = \max\big(0, (B_{ij,l}B_{ij,l}^{T}+I)^{-1}(B_{ij,l}(\lambda \bs{d}_{ij} + \bs{z}_{ij,0}^{(k+1)} + \mu \bs{y}_{ij}^{(k)}-\sum_{p\neq l}B_{ij,p}^{T}\bs{z}_{ij,1,p}^{\star})+\bs{z}_{ij,2,l}^{(k)})\big). \nonumber \
%\bs{z}_{ij,1,2}^{\star} = \max\big(0, (B_{ij,2}B_{ij,2}^{T}+I)^{-1}(B_{ij,2}(\lambda \bs{d}_{ij} + \bs{z}_{ij,0}^{(k+1)} + \mu \bs{y}_{ij}^{(k)}-B_{ij,1}^{T}\bs{z}_{ij,1,1}^{\star})+\bs{z}_{ij,2,2}^{(k)}\big). \nonumber
\end{align}
where $B_{ij,l}B_{ij,l}^{T}+I$ is a diagonal matrix and $\bs{z}_{ij,1,p}^{\star}$ is the latest value of $\bs{z}_{ij,1,p}$.

\para{Step 3: Optimizing $\bs{z}_{ij,2}^{(k+1)}$, $\bs{z}_{ij,3}$, $\bs{z}_{ij,4}$, $Z_{ij,5}$ and $Z_{ii,6}$.} The optimization of each $\bs{z}_{ij,2}$ is decoupled and its optimal value is
\begin{align}
\bs{z}_{ij,2}^{(k+1)} & = \arg\min_{\bs{z}_{ij,2}}\|\set{F}^{T}(\bs{z}_{ij,2}) - \Big(\bs{z}_{ij,3}^{(k)}\bs{1}^{T}+\bs{1}{\bs{z}_{ij,4}^{(k)}}^{T}-C-Z_{ij,5}^{(k)}  - 2S_{ij}^{(k)}-\mu X_{ij}^{(k)}\Big)\|_{F}^2 + \|\bs{z}_{ij,2} - \bs{z}_{ij,1}^{(k+1)}\|^2 \nonumber \\
& = (\set{F}\set{F}^{T}+I)^{-1}\Big(\set{F}\big(\bs{z}_{ij,3}^{(k)}\bs{1}^{T}+\bs{1}{\bs{z}_{ij,4}^{(k)}}^{T}-C-Z_{ij,5}^{(k)}  - 2S_{ij}^{(k)}-\mu X_{ij}^{(k)}\big) + \bs{z}_{ij,1}^{(k+1)}\Big).
\end{align}
where $\set{F}\set{F}^{T}+I$ is a diagonal matrix.
%, so $\bs{z}_{ij,2}^{(k+1)}$ can be efficiently computed.
Through a similar derivation, we can compute the optimal value of other variables at iteration $k+1$ as
\begin{align}
\bs{z}_{ij,3}^{(k+1)} &= \arg\min_{\bs{z}_{ij,3} \geq 0} \|\bs{z}_{ij,3}\bs{1}^{T} -\Big(C_{ij} + \set{F}^{T}(\bs{z}_{ij,2}^{(k+1)})+Z_{ij,5}^{(k)}+\mu X_{ij}^{(k)} - \bs{1}{\bs{z}_{ij,4}}^{(k)}\Big)\|_{F}^2 + 2\mu \langle \bs{1}, \bs{z}_{ij,3}\rangle \nonumber \\
&= \max(0, \Big(\big(C_{ij} + \set{F}^{T}(\bs{z}_{ij,2}^{(k+1)})+Z_{ij,5}^{(k)}+\mu X_{ij}^{(k)} - \bs{1}{\bs{z}_{ij,4}}^{(k)}\big)\bs{1} - \mu\Big)/|V_j|)
\end{align}

\begin{align}
\bs{z}_{ij,4}^{(k+1)} &= \arg\min_{\bs{z}_{ij,4} \geq 0} \|\bs{1}\bs{z}_{ij,4}^{T} -\Big(C_{ij} + \set{F}^{T}(\bs{z}_{ij,2}^{(k+1)})+Z_{ij,5}^{(k)}+\mu X_{ij}^{(k)} - {\bs{z}_{ij,3}}^{(k+1)}\bs{1}^{T}\Big)\|_{F}^2 + 2\mu \langle \bs{1}, \bs{z}_{ij,4}\rangle \nonumber \\
&= \max(0, \Big(\big(C_{ij} + \set{F}^{T}(\bs{z}_{ij,2}^{(k+1)})+Z_{ij,5}^{(k)}+\mu X_{ij}^{(k)} - \bs{1}{\bs{z}_{ij,4}}^{(k)}\big)^{T}\bs{1} - \mu\Big)/|V_i|)
\end{align}

\begin{align}
Z_{ij,5}^{(k+1)} & = \arg\min_{Z_{ij,5} \geq 0}\|Z_{ij,5} - \Big(\bs{z}_{ij,3}^{(k+1)}\bs{1}^{T}+\bs{1}{\bs{z}_{ij,4}^{(k+1)}}^{T} - C_{ij}- \set{F}_2^{T}(\bs{z}_{ij,2}^{(k+1)})-\mu X_{ij}^{(k)} - 2S_{ij}^{(k)}\Big)\|_{F}^2 \nonumber \\
& = \max\Big(0,\bs{z}_{ij,3}^{(k+1)}\bs{1}^{T}+\bs{1}{\bs{z}_{ij,4}^{(k+1)}}^{T} - C_{ij}- \set{F}_2^{T}(\bs{z}_{ij,2}^{(k+1)})-\mu X_{ij}^{(k)} - 2S_{ij}^{(k)} \Big), \nonumber \\
Z_{ii,6}^{(k+1)} & = \arg\min_{Z_{ii,6} \geq 0}\|S_{ii} + \mu X_{ii} - Z_{ii,6}-\mu I_{|V_i|}\|_{F}^2\nonumber \\
& = \max\Big(0,S_{ii}^{(k)} + \mu X_{ii} ^{(k)}- \mu I_{|V_i|}\Big).
\end{align}

\para{Step 4: Optimizing $S$.} Finally, we optimize $S$. In this case, the optimization problem is reduced to
\begin{align}
S^{(k+1)} & = \arg\min_{S \geq 0}\sum\limits_{1\leq i < j \leq N}\|C_{ij} + \mathcal{F}_{ij}^{T}(\bs{z}_{ij,2}^{(k+1)})- \bs{z}_{ij,3}^{(k+1)}\bs{1}^{T} - \bs{1}{\bs{z}_{ij,4}^{(k+1)}}^{T} + Z_{ij,5}^{(k+1)} + 2S_{ij} + \mu X_{ij}^{(k)}\|_{F}^2 \nonumber \\
 & \qquad\quad +  \sum\limits_{1\leq i \leq N}\|S_{ii}- Z_{ii,6}^{(k+1)} + \mu X_{ii}^{(k)}\|\\
& = \arg\min_{S \geq 0}\|S - T^{(k)}\|_{F}^2 \nonumber \\
& = U\max(\Sigma, 0)U^{T},
\end{align}
%\textcolor{red}{Please make sure the below equation is correct. Marked by Jinbo Xu}
where $U\Sigma U^{T}$ is eigen-decomposition of $T^{(k)}$ defined below and $\max(\Sigma, 0)$ takes the positive eigenvalues.\begin{align}
T_{ij}^{(k+1)} = \left\{
\begin{array}{cc}
\Big(\bs{z}_{ij,3}^{(k+1)}\bs{1}^{T}+\bs{1}{\bs{z}_{ij,4}^{(k+1)}}^{T} - C_{ij} - F_{ij}^{T}(\bs{z}_{ij,2}^{(k+1)}-\mu X_{ij}^{(k)}\Big)/2& i \neq j\\
Z_{ii,6}^{(k+1)}-\mu X_{ii}^{(k)} & \textup{otherwise}
\end{array}
\right.\
\end{align}

\para{Step 5: Optimizing primal variables.} Finally the primal variables are updated as follows:
\begin{align}
\bs{y}_{ij}^{(k+1)} &= \bs{y}_{ij}^{(k)} + \frac{1}{\mu}\Big(\lambda \bs{1} + \bs{z}_{ij,0}^{(k+1)} - B^{T}\bs{z}_{ij,1}^{(k+1)} \Big), \nonumber \\
X_{ij}^{(k+1)} &= X_{ij}^{(k)} + \frac{1}{\mu}\Big(C_{ij} + \set{F}_{ij}^{T}(\bs{z}_{ij,2}^{(k+1)}) - \bs{z}_{ij,3}^{(k+1)}\bs{1}^{T} - \bs{1}{\bs{z}_{ij,4}^{(k+1)}}^{T} + Z_{ij,5}^{(k+1)}+2S_{ij}^{(k+1)}\Big) \nonumber \\
\bs{s}^{(k+1)} &= \bs{s}^{(k)} + \frac{1}{\mu}(\bs{z}_1^{(k+1)} - \bs{z}_2^{(k+1)}).
\end{align}

\end{document}